\documentclass[aps,prb,showpacs,preprint,preprintnumbers,amsmath,amssymb]{revtex4}

\usepackage{amssymb}
\usepackage{amsmath}
\usepackage{graphicx}

\begin{document}
\title{Berry curvature mediated valley Hall and charge Hall effects in graphene via strain engineering}

\author{Zhen-Gang Zhu and Jamal Berakdar}
\affiliation{Institut f\"ur Physik, Martin-Luther-Universit\"at Halle-Wittenberg, 06099 Halle, Germany}

\begin{abstract}
{ We point out a practical way to induce a finite charge Hall conductivity in graphene via an off-diagonal strain.
 Our conclusions are based on a general analysis and supported by
numerical examples.
The interplay between a substrate-induced   gap and the strain is discussed. We show that the substrate-induced gap itself does not break the symmetry that maps one valley onto the other, and hence does not yield a net charge Hall effect.  The net charge Hall conductivity is found to be sizable and should be  observable   in the presence of both the gap and the strain. The sign of the Hall conductivity
can be tuned by varying the strain parameters. The valley Hall conductivity is studied as well. The valley and the orbital magnetic moment accumulations on the sample edges may be detected in a  Hall conductivity experiment.}
\end{abstract}


\maketitle
\section{Introduction}
 The hallmark of graphene, a single layer of carbon atoms arranged on a honeycomb lattice, is its particular band structure with a  linear dispersion of the low-energy carriers
  around the two inequivalent (Dirac) points $K$ and $K'$ at the corners of the first Brillouin zone \cite{graphene}.
  The $K$ and $K'$ valleys are distinguished by the valley (isospin) degree of freedom which is
  robust against perturbations (unless a  momentum on the scale of the graphene's reciprocal lattice vector is transferred) \cite{foot},
   which makes  it attractive for (valleytronics) applications  \cite{1,2,xiao07,4,5,6}.
   Hence, it is highly desirable to explore whether transport properties are controllable
     via the valley degree of freedom.
        As well-established \cite{berry}, the charge Hall conductivity  $\sigma^{\text{C}}_{xy}$ is governed  by  the Berry phase, which for each Dirac cone is finite ($\pm\pi$ for $K(K')$), but the sum is zero. The Berry curvature mediated Hall effect in graphene
        has been  studied  in the presence of a spin-orbit coupling \cite{kane05,qiao10,tse11}, or for a substrate-induced gap \cite{xiao07,zhang09,xiao10,fzhang11}, or in the proximity  to a superconductor \cite{ghaemi}.
   For a monolayer graphene with a uniform interaction gap, e.g. as resulting from  the interaction with a substrate,
one finds   \cite{xiao07} a finite Berry curvature and a  finite charge Hall conductivity $\sigma^{K(K')}_{xy}$ in each of the Dirac cones by the interaction gap \cite{fuchs}. The total charge Hall current vanishes however, for the  Berry curvature and the charge Hall conductivity for the two Dirac cones are equal with opposite signs. To circumvent this problem and obtain a net charge Hall current the authors of Ref. \cite{xiao07} proposed the creation of   a non-equilibrium population in the  Dirac cones, which is inherently hard to realize and sustain in a real device.

{An energy gap in the electronic structure  is advantageous for  graphene applications. E.g., a zero gap in graphene prevents the graphene field-effect transistor (GFET) device from possessing  a turn-off state \cite{gfet}. A substrate-induced gap is not  restricted to the SiC substrate but is also observed for the hexagonal boron nitride substrate \cite{bn}. A further  way to tune the energy gap is via the strain engineering, e.g. as
brought about by strain super lattices or wrinkles \cite{low2011}. The strain can be viewed  as a symmetry-adapted effective, pseudo magnetic field in the sublattice space for the two valleys \cite{low,vozmediano,fujita10}. A uniaxial strain can be realized  experimentally  by bending a flexible substrate \cite{mohiuddin}. And a biaxial strain in graphene can be created by shallow depressions \cite{metzger}. Recently, a zero-field quantum Hall effect in graphene was proposed via a designed strain profile \cite{guinea2010}. Usually, strain effect is studied via a minimal coupling which is momentum independent \cite{low2011}.
Here we go beyond the minimal coupling and include  terms that are linear in the momentum by using the theory of invariants \cite{winkler}. The reason is that these terms will change the topology of the Dirac cones and which is manifested in real physical quantities, for example the Hall conductivity. Furthermore, the effect of strain is usually thought to introduce time-reversal-invariant terms in the Hamiltonian. However, we show in a time-reversal-breaking environment, for example under a weak magnetic field, the strain could couple with graphene via a time-reversal-breaking term which gives rise to exotic effects,
   e.g., under the strain,  a net charge Hall current can be driven by an applied electric field. 
   This effect may be useful to design strained-graphene-based devices where the Hall current can be turned on and off via an in-plane weak magnetic field.}

\section{Model}  For a general analysis, let us start from a Hamiltonian derived by the theory of invariants that accounts systematically for all possible strains  \cite{winkler}. Symmetry operations 
and their compatibility with the underlying lattice
structures are encapsulated in this approach.

The general structure of the two-dimensional (2D) Dirac Hamiltonian  is
\begin{eqnarray}
h_{K}(\mathbf{k}) &=& d_{0}(\mathbf{k})+\boldsymbol{\sigma}\cdot\mathbf{d}(\mathbf{k}), \text{ for   $K$ cone,} \notag\\
h_{K'}(\mathbf{k}) &=& {d}_{0}'(\mathbf{k})+\boldsymbol{\sigma}\cdot\mathbf{d}'(\mathbf{k}),  \text{ for  $K'$ cone.}
\label{hkkp}
\end{eqnarray}
The vectors $\mathbf {d}(\mathbf{k})$ and $\mathbf {d}'(\mathbf{k})$ map the 2D momentum space to a 3D parameter space. $d_0(\mathbf{k})$ and $d'_{0}(\mathbf{k})$ are structureless parameters in the sublattice space and yield no contributions to the Berry curvature and Hall effect (see the appendix for details). The concrete form of these vectors depends on the realized physical situations  and is given and discussed  in full details  in the appendix.
$\mathbf{d}$ is parameterizable as $\mathbf{d}=d(\cos\varphi\sin\theta,\sin\varphi\sin\theta,\cos\theta)$ (for  $\mathbf{d'}$, we use $d'$, $\varphi'$ and $\theta'$).  The eigen energies are  $E_{\pm}=d_{0}(\mathbf{k})\pm d(\mathbf{k})$ and $E'_{\pm}=d'_{0}(\mathbf{k})\pm d'(\mathbf{k})$. Solving for the  eigenfunctions of Eqs. (\ref{hkkp}), we can obtain the Berry connection (i.e. the fictitious vector potential as explained in the appendix) and the Berry curvature (i.e. the $z$ component of the associated magnetic field in the momentum space)  as
\begin{equation}
\Omega_{\nu Kz}(\mathbf{k})=\nu\frac{\sin\theta}{2}\left(\frac{\partial\theta}{\partial k_{x}}\frac{\partial\varphi}{\partial k_{y}}-\frac{\partial\theta}{\partial k_{y}}\frac{\partial\varphi}{\partial k_{x}}\right),
\label{berrycurvature}
\end{equation}
where $\nu=+(-)$ for conduction (valence) band. The same applies for $\Omega_{\nu K'z}$, with $\theta$ and $\varphi$
 being replaced by $\theta'$ and $\varphi'$. The charge current operator is $j_{i}=\frac{\partial h(\mathbf{k})}{\partial k_{i}}$. Introducing $\tau_{z}$ as the Pauli matrix along the $z$ direction to describe the valley degree of freedom
  we write for the valley current operator  $j^{\text{V}}_{i}=\frac{1}{2}\{\tau_{z},j_{i}\}$, where $i=x,y$.  From the standard Kubo formula \cite{qi2006} given in the appendix, we find the charge Hall ($\sigma^{\text{C}}_{xy}$)  and the valley Hall ($\sigma^{\text{V}}_{xy}$) conductivities   as
\begin{eqnarray}
\sigma^{\text{C(V)}}_{xy} &=& \sigma_{xy}^{K}\pm\sigma_{xy}^{K'} \label{hallconductivity} \\
&=& \frac{e^{2}}{\hbar A_{0}}\left[\sum_{\nu\mathbf{k}}n_{\nu}(\mathbf{k})\Omega_{\nu Kz}(\mathbf{k})\pm\sum_{\nu\mathbf{k'}}n_{\nu}(\mathbf{k'})\Omega_{\nu K'z}(\mathbf{k'})\right]. \notag
\end{eqnarray}
where  $A_{0}$ is the area of the system, $n_{\nu}(\mathbf{k})$ is the Fermi function.
The Berry curvature can be further expressed as
\begin{equation}
\Omega_{\nu Kz}(\mathbf{k})=\frac{\nu}{2d^{2}(\mathbf{k})}\varepsilon_{\alpha\beta\gamma}j_{\alpha x}j_{\beta y}\hat{d}_{\gamma},
\label{berry2}
\end{equation}
where $\varepsilon_{\alpha\beta\gamma}$ is the anti-symmetric tensor, $j_{\alpha i}=\frac{\partial d_{\alpha}(\mathbf{k})}{\partial k_{i}}$ is the \textit{pesudo spin current tensors}, $\alpha,\beta,\gamma=1,2,3$ and $\hat{d}_{\gamma}=d_{\gamma}/d(\mathbf{k})$.
\begin{figure}[tph]
\centering
\includegraphics[width=0.8\textwidth]{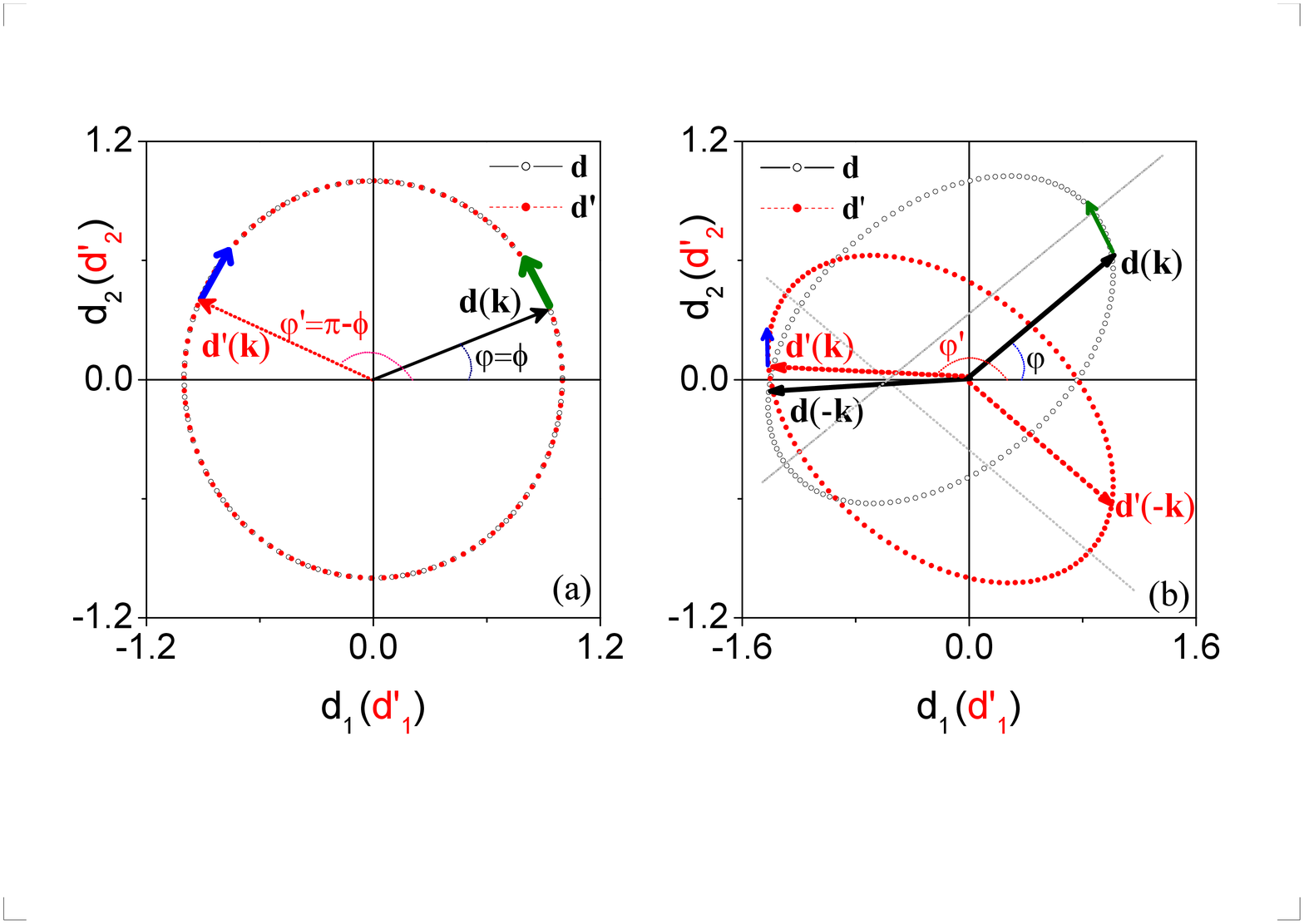}
\caption{(color online) The \textit{d} vector space which is mapped from a circle in the momentum space is shown for the case in absence (a)  or presence (b) of strain. The unit of d vector is $\tilde{v}_{F}k$. The other parameters in (b) are $\frac{\tilde{v}_{x}}{\tilde{v}_{F}}=\frac{\tilde{v}_{y}}{\tilde{v}_{F}}=\frac{\mathcal{A}_{x}}{\tilde{v}_{F}k}=\frac{\mathcal{A}_{y}}{\tilde{v}_{F}k}=0.2$.
\label{d1d2}}
\end{figure}

\section{Hamiltonian with strain and analysis}
\subsection{Analysis for the case without strain}
In absence of the strain and the interaction gap the vector $\mathbf{d(d')}$ has only in-plane components, i.e. $d_{1}(d'_{1})=\pm v_{F}k_{x}$ and $d_{2}(d'_{2})=v_{F}k_{y}$. Geometrically, a circle in momentum space (constant $|\mathbf{k}|$) is mapped onto a circle in the \textit{d} vector space shown in Fig. \ref{d1d2}(a). The $\mathbf{d}$ and $\mathbf{d'}$ are related by a reflection  $R^{-1}_{y}$ (a reflection through $yz$ plane which is perpendicular to the graphene layer see \cite{winkler}).
This operation conserves the sublattices but exchanges the Dirac cones. Under this operation, $(k_{x},k_{y})$ in $K$ cone is
 transformed to  $(-k_{x},k_{y})$ in $K'$ cone which means the angle $\phi$ of $\mathbf{k}$ (i.e. $\tan\phi=k_{y}/k_{x}$) is mapped
 upon  $R^{-1}_{y}$ to an angle $\pi-\phi$ of $(-k_{x},k_{y})$ in $K'$ cone. The role of $R^{-1}_{y}$ in the \textit{d} vector space is manifested as a reflection through $d_{2}d_{3}$ plane. Therefore, we have $\varphi=\phi$, $\varphi'=\pi-\phi$ and $\theta'=\theta$ between the phases of  $\mathbf{d}$ and $\mathbf{d}'$. Thus, with increasing $\phi$ in $K$ cone ($\mathbf{d}$), the reflected vector in $K'$ cone ($\mathbf{d}'$) varies in the opposite direction which explains the cancelation of the total charge Hall current (see Fig. \ref{d1d2}(a)). The above analysis is also valid in the presence of an interaction gap although a $d(d')_{3}$ is introduced and a nontrivial topology in each Dirac cone is induced by it. The interaction gap does not break the symmetry between the two Dirac cones \cite{xiao07}.

\subsection{Analysis for the case with strain}
To discuss the effect of the strain, let us first define the strain tensor as
\begin{equation}
u_{ij}=\frac{1}{2}\left(\frac{\partial u_{i}}{\partial r_{j}}+\frac{\partial u_{j}}{\partial r_{i}}+\frac{\partial u_{z}}{\partial r_{i}}\frac{\partial u_{z}}{\partial r_{j}}\right),
\label{straintensor}
\end{equation}
where $\mathbf{u}$ is a polar vector indicating the displacements of atoms. In the theory of invariants \cite{toi}, the strain tensor couples with $\mathbf{k}$ giving rise to the irreducible tensors $\mathcal{K}$ under $D_{3h}$ group. We only focus on the $H^{55}$ which is derived by  a multiplication of the irreducible matrices and the irreducible tensors that belong both to the $\Gamma_{5}$ representations. Thus, the Hamiltonian is expressed in terms of $\Gamma_{1}$, $\Gamma_{2}$ and $\Gamma_{6}$, respectively. The corresponding irreducible matrices for these representations are $\mathbf{1}$, $\sigma_{z}$, and $(\sigma_{x},\sigma_{y})$ \cite{winkler}. Hence, we infer $d_{0}=\mathcal{K}^{1}$, $d_{3}=\mathcal{K}^{2}$, $d_{1}=\mathcal{K}^{6,1}$ and $d_{2}=\mathcal{K}^{6,2}$ where $\mathcal{K}^{\kappa,\lambda}$ indicates irreducible tensors that transform according to the  $\kappa$-th irreducible representations of $D_{3h}$ \cite{winkler}.

From the theory of invariants \cite{winkler} (cf. also the appendix), we obtain the $\mathbf{d}$ vector in the $K$ and $K'$ cones {without including the external electric field}
\begin{eqnarray}
d_{1}&=& (\tilde{v}_{F}k_{x}+\tilde{\mathbf{v}}\cdot\mathbf{k})\tau_{z}-\mathcal{A}_{x}, \notag\\
d_{2} &=& \tilde{v}_{F}k_{y}+(\mathbf{k}\times\tilde{\mathbf{v}})_{z}+\mathcal{A}_{y}\tau_{z}, \notag\\
d_{3} &=& \beta(\tilde{v}_{x}k_{y}+\tilde{v}_{y}k_{x}),
\label{dvector1}
\end{eqnarray}
where $\tau_{z}$ denotes the valley degree, $\tilde{v}_{F}=v_{F}+b_{62}(u_{xx}+u_{yy})$, $\tilde{v}_{x}=b_{63}(u_{xx}-u_{yy})$, $\tilde{v}_{y}=2b_{63}u_{xy}$, and $\beta=b_{21}/b_{63}$ being a dimensionless parameter.
As in studies  on semiconductors  all the  parameters  ($b$'s) are to be determined experimentally
 or from \textit{ab-initio} calculations \cite{toi}.
  The effective vector potential whose components are $\mathcal{A}_{x}=b_{61}(u_{xx}-u_{yy})$ and $\mathcal{A}_{y}=b_{61}(2u_{xy})$ was studied as a minimal coupling in a strain-induced transport \cite{pereira,pereira2009,neto}. This vector potential can also be expressed in terms of the hopping parameters in a tight-binding formalism $\mathcal{A}_{x}=\frac{\sqrt{3}}{2}(t_{3}-t_{2})$ and $\mathcal{A}_{y}=\frac{1}{2}(t_{2}+t_{3}-2t_{1})$ where $t$'s are the hopping parameters under strain \cite{vozmediano}. This term is independent of $\mathbf{k}$ and can not change the topology of the Dirac cones. {Please note the third component, i.e. $d_{3}$, is present under the time-reversal-breaking environment. This term violates the $\mathbf{d}$ vector from $\theta=\pi/2$ plane. Thus the $\mathbf{d}$ vector now describes a map from a 2D momentum space to a 3D vector space. We will study the specific  consequences of this mapping.}
Furthermore, the diagonal strain $u_{xx}-u_{yy}$ and the off-diagonal component $u_{xy}$ yield new effects,  beyond the minimal coupling, and  give rise to extra terms that depend on $\mathbf{k}$. Therefore, the topology of the cones is changed and a non-vanishing Hall conductivity will be the first-order effect from these terms.

Note, that the $b_{62}$ term in the above equations stems from the contribution of a smooth rippling of the graphene sheet \cite{winkler,juan} and results
 in  an isotropic renormalization of the electron velocity. The action of the operator $R^{-1}_{y}$ is now $R^{-1}_{y}(\mathcal{A}_{x},\mathcal{A}_{y})=(\mathcal{A}_{x},-\mathcal{A}_{y})$, $R^{-1}_{y}(\tilde{v}_{x},\tilde{v}_{y})=(\tilde{v}_{x},-\tilde{v}_{y})$.
Because of these symmetry relations, in the presence of a strain, a circle in the momentum space is mapped onto  ellipses in the \textit{d} vector space, as  visualized in Fig. \ref{d1d2}(b). 
The two ellipses are reflection-symmetric  at $d_{1}d_{3}$ plane in the \textit{d} vector space. This plane is perpendicular to the one in the case of zero strain. This is the consequence of the symmetry property for the strain tensor $u_{ij}$ under $R_{y}^{-1}$. $u_{ij}$ is not a polar vector (say $\mathbf{k}$) and transforms like the symmetrized $\left\{k_{i},k_{j}\right\}$. However, the symmetry plane of $d_{1}d_{3}$ links $\mathbf{d}(\mathbf{k})$ and $\mathbf{d}'(-\mathbf{k})$ and vice versa (rather then  $\mathbf{d}(\mathbf{k})$ and $\mathbf{d}'(\mathbf{k})$) (see  Fig. \ref{d1d2}(b)). We note $d_{3}(\mathbf{k})=-d'_{3}(-\mathbf{k})$. Therefore, we can rearrange the summation in deriving  $\sigma^{\text{C(V)}}_{xy}$ over $\mathbf{k}'$ in $K'$ cone by the restriction $\mathbf{k}'=-\mathbf{k}$ in Eq. (\ref{hallconductivity}). So the phases of $d'$ are $\phi'=2\pi-\phi$ and $\theta'=\pi-\theta$. According to  Eq. (\ref{berrycurvature}) we obtain the remarkable result
\begin{equation}
\Omega_{\nu Kz}=\Omega_{\nu K'z}.
\label{berrystrain}
\end{equation}
 The two Dirac cones have the same conductivity contributions  in contrast to the opposite contributions  in zero strain case.
 The opposite  properties  of the Berry curvature for the conduction  and the valence bands are  maintained since the strain does not break the particle-hole symmetry. Eq. (\ref{berrystrain}) evidences a net charge Hall conductivity in our case.

\section{Numerical illustrations}
\begin{figure}[tphb]
\centering
\includegraphics[width=0.8\textwidth]{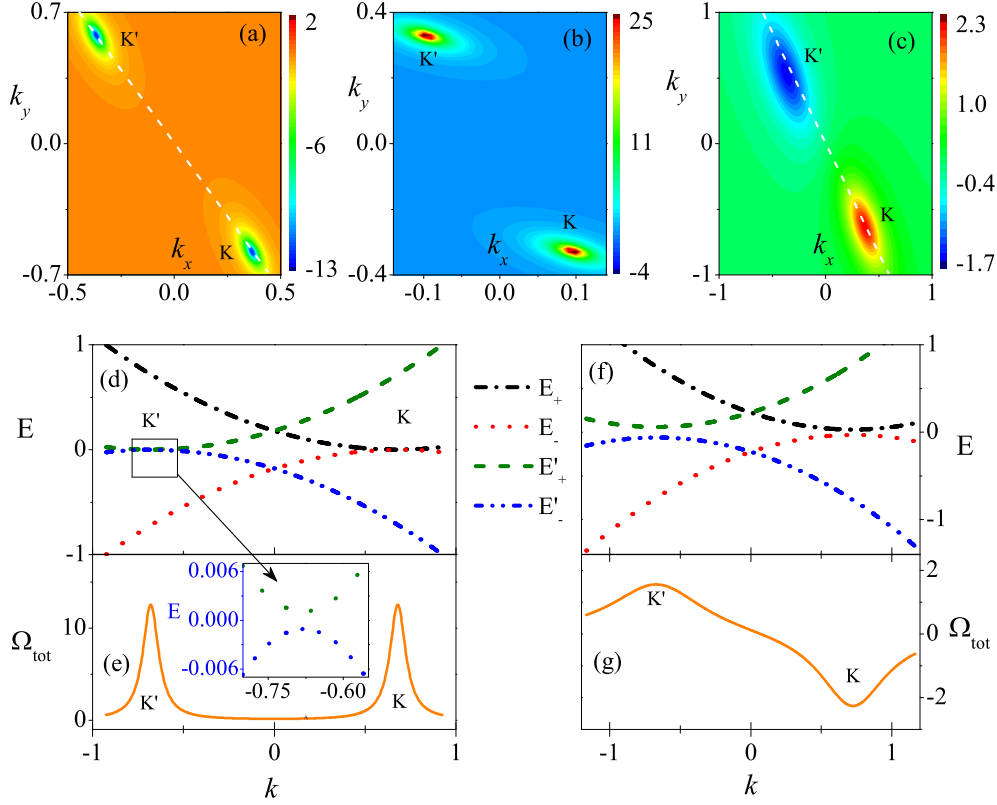}
\caption{(color online) The contour plot of the total Berry curvature $\Omega_{\text{tot}}$ for the conduction band is calculated numerically in (a), (b) and (c). The energy profiles in (d) and (f) and the $\Omega_{\text{tot}}$ in (e) and (g) are calculated along the dashed lines in (a) and (c) respectively. The inset of (d) and (e) is a zoom-in plot  around  $K'$ cone. The units are $k_{0}=1$ nm$^{-1}$, $\tilde{v}_{F}$, $\tilde{v}_{F}\hbar k_{0}$ for the wave vector, the velocity, and energy respectively (note the unit of $\mathcal{A}$ is energy). 
$\beta=0.5$ is kept for all graphs. A energy gap $\Delta=0.28$ eV \cite{zhou} is present in (c), (f) and (g) and zero elsewhere. The other parameters are $\frac{\tilde{v}_{x}}{\tilde{v}_{F}}=\frac{\mathcal{A}_{x}}{\tilde{v}_{F}k_{0}}=\frac{\tilde{v}_{y}}{\tilde{v}_{F}}=\frac{\mathcal{A}_{y}}{\tilde{v}_{F}k_{0}}=0.3$ in (a), (c)-(g); $\frac{\tilde{v}_{x}}{\tilde{v}_{F}}=\frac{\mathcal{A}_{x}}{\tilde{v}_{F}k_{0}}=0$, $\frac{\tilde{v}_{y}}{\tilde{v}_{F}}=\frac{\mathcal{A}_{y}}{\tilde{v}_{F}k_{0}}=0.3$ in (b).
\label{berrycur}}
\end{figure}

To demonstrate the symmetry analysis, in Fig. \ref{berrycur} we show the results of the numerical calculations for  the total Berry curvature in (a), (b) and (c). The observations from (a) and (b) are:\\
\emph{ i)} the prominent peaks (or dips) indicate the positions of the Dirac points in  momentum space
  shifted by the effective magnetic field induced by the strain.\\
  \emph{ii)} A singularity of these peaks (or dips) does  not exist generally since a strain-induced gap is present.\\
   \emph{iii)} The Berry curvature apart from the Dirac points is generally finite.\\
  \emph{  iv) } The two peaks in  Berry's curvature are just at the antipodal points with respect to the zero $\mathbf{k}$ (inversion in the momentum space).\\
 \emph{  v)} The most important fact is that the  Berry curvatures at these two points possess the same sign.
  These observations are exactly what one  expects  from a symmetry analysis. Furthermore, we  find  the Berry curvature is exactly zero if the off-diagonal component $u_{xy}$ is zero. For  $u_{xy}\neq 0$,  as shown in Fig. \ref{berrycur}(b), the  Berry curvature is non vanishing even when $u_{xx}=u_{yy}$ (which means $\mathcal{A}_{x}=\tilde{v}_{x}=0$).
  Note, the discussed symmetry is still preserved with the result of a
  net Hall conductivity.
  The difference to Fig. \ref{berrycur}(a) and (b) is that the Hall conductivity is present with a reversed sign. The energies (with respect to $d_{0}(d'_{0})$) for the conduction  and valence bands,  and the total Berry curvature  calculated along the dash line in (a), are shown in Fig. \ref{berrycur}(d) and (e), respectively. As evident, the two peaks in the Berry curvature are the same and appear at the positions where the smallest band gap exists.
  The inset in (e) shows a zoom-in  of the opened  strain-induced gap.

In Fig. \ref{berrycur}(c) shows the calculated the total Berry curvature for the conduction band in the presence of a strain and an interaction gap \cite{zhou}. The interaction gap is uniformly generated for both Dirac cones \cite{xiao07} and thus adds to  $d_{3}$ and $d'_{3}$. The interaction gap breaks the symmetry shown in Fig. \ref{d1d2}(b). If $d_{3}=d'_{3}=\Delta/2$ only, the odd-symmetry of the Berry curvature  for the two cones is preserved. If there is only a strain effect in  $d_{3}$ and $d'_{3}$, the symmetry is changed to  even. If the interaction gap and the strain  are both present, they compete with each other. Generally,  the Berry curvature does not possess an explicit symmetry which can be seen from  Figs. \ref{berrycur}(f) and (g) along the dash line in (c). The gaps at the shifted $K$ and $K'$ cones composed of the interaction gap and the  strain-induced gap  are generally not symmetric now.

\begin{figure}[tphb]
\centering
\includegraphics[width=0.8\textwidth]{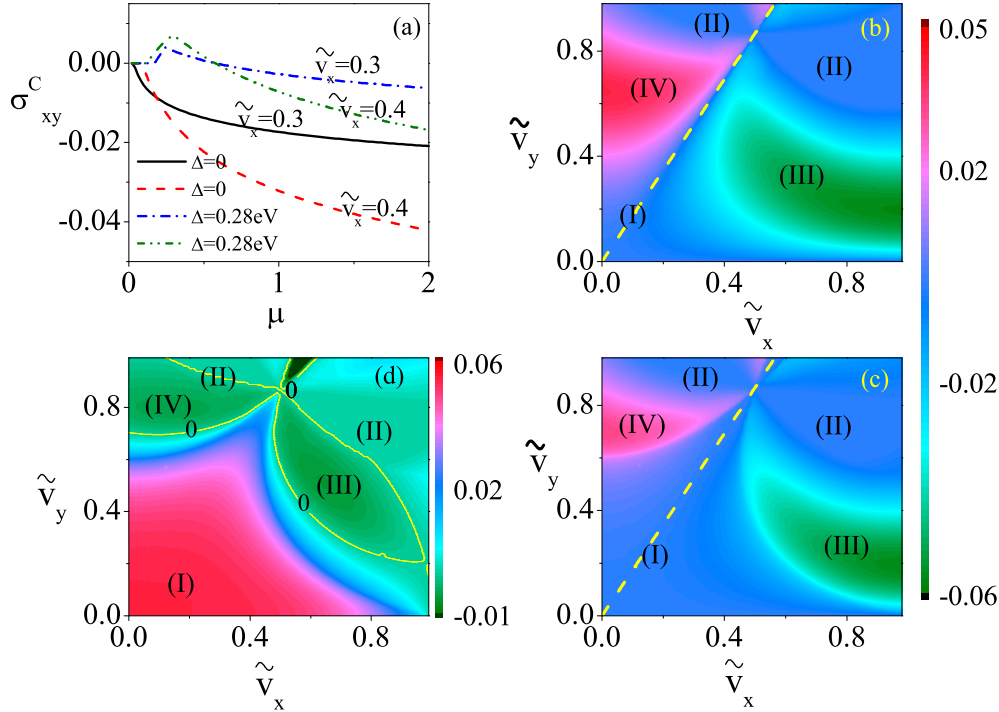}
\caption{(color online) The dependencies  of  $\sigma^{\text{C}}_{xy}$ on $\mu$ in (a)  and on  $\tilde{v}_{x}$  and $\tilde{v}_{y}$ in (b) and (c). (d) shows the variation of $\sigma^{\text{V}}_{xy}$ with $\tilde{v}_{x}$ and $\tilde{v}_{y}$. In (a), $\tilde{v}_{x}=\tilde{v}_{y}$, $\alpha=1$ and $\beta=0.5$. 
The parameters in (b), (c) and (d) are $\beta=0.5$, $\alpha=1.2$ and $\mu=1.0$. The unit of the conductivity is $e^{2}/\hbar$. The other units are the same as in Fig. \ref{berrycur}.
\label{hallstrain}}
\end{figure}

Introducing the relation $\mathcal{A}_{x(y)}=\alpha\tilde{v}_{x(y)}$ we calculated the total charge Hall conductivity with varying the Fermi level  $\mu$ (Fig. \ref{hallstrain} (a)). For graphene without the interaction gap there is no sign change in $\sigma^{\text{C}}_{xy}$. 
In the presence of the interaction gap and the strain,  there is a sign change since the interaction gap is modified to two non-symmetric gaps in the two cones (see Fig. \ref{berrycur} (c) and (f)). With increasing $\mu$, one cone dominates first. The other cone follows up with a further increase of $\mu$ leading thus to a sign change. The values at which the sign change occurs indicate the asymmetry  of the gaps at $K$ and $K'$.
 The saturation value of the $\sigma^{\text{C}}_{xy}$ is increased with a larger strain. To obtain a deeper insight into how the $\sigma^{\text{C(V)}}_{xy}$ varies with the strain parameters, the contour plots are shown in Fig. \ref{hallstrain}(b)-(d). In (b) and (c), $\sigma^{\text{C}}_{xy}$ is shown for $\Delta=0$ and $\Delta=0.28$ eV, respectively.
The  regions of a finite Hall conductivity are split  into islands by a zero regions. The dash lines in (b) and (c) indicate the relation $\tilde{v}_{y}=\sqrt{3}\tilde{v}_{x}$ from which $d(d')=0$ can be obtained. When this happens there is no structure in the system. In the upper part of this line, the islands show a positive sign of the  charge Hall conductivity.  In contrast, in the lower part of the line, the charge Hall conductivity is negative. 
In regions (III) and (IV) in Figs. \ref{hallstrain}(b) and (c), the strain effect is prominent leading to a large  charge Hall conductivity.  The areas of the regions (III) and (IV) shrink accordingly in  Fig. \ref{hallstrain}(c) with nonzero interaction gap.
An interesting observation is that a transition may occur when  crossing from the lower part to the upper part. Noting that the strain is coupled with the graphene in a symmetry-governed way, the symmetry at the transition point is maintained and the topology is changed. This could be an example of a Lifshitz transition \cite{volovik} induced by the strain in graphene. In the regions (I) in Figs. \ref{hallstrain}(b) and (c), $\sigma^{\text{C}}_{xy}$ is small since the interaction gap dominates the  strain-induced gap  in (c) and a small  strain-induced gap  in (b). 
The regions (II) in (b) and (c) evidence that the induced gap is  large  leading to a zero $\sigma^{\text{C}}_{xy}$ for each cone.
The sign change of $\sigma^{\text{C}}_{xy}$ is caused by the reversed sign of the cross products of the two pseudo spin currents, i.e. $\mathbf{j}_{x(y)}=(j_{1x(y)},j_{2x(y)},j_{3x(y)})$ in 3D $\hat{d}$ space.
 Fig. \ref{hallstrain}(d) shows the valley Hall conductivity . In contrast to the corresponding regions in (b) and (c), the $\sigma^{\text{V}}_{xy}$ is large in region (I) and possesses the same sign for the (III) and (IV) regions. In regions (II), the valley Hall conductivity  is quite small as well.

\begin{figure}[tphb]
\centering
\includegraphics[width=0.5\textwidth]{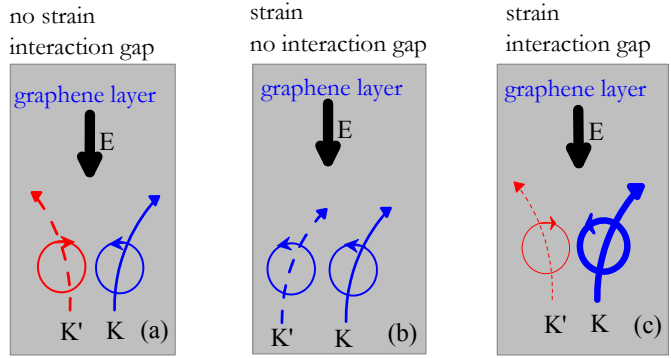} 
\caption{(color online) Schematics  of the charge and valley Hall effect under an external electric field  for   a zero strain and nonzero interaction gap in (a), nonzero strain and zero interaction gap in (b), and nonzero strain and nonzero interaction gap in (c). The arrowed curves denote the electron flows. The arrowed circles indicate the orbital magnetic momentum in $K$ and $K'$ cones. The thicker (thinner) curve means larger (weaker) electron current.
\label{CHCVHC}}
\end{figure}
\section{Possible experimental setup}  Experimentally,  the valley degree could be characterized by the valley-carried orbital magnetic moment \cite{xiao07,xiao10,fuchs}. In our study, the orbital magnetic moment in the presence of the strain is $m^{K(K')}_{\nu z}(\mathbf{k})=-\nu\frac{ed(d')}{\hbar}\Omega_{\nu K(K')z}(\mathbf{k})$, where $\nu=\pm$ for the conduction and the valence band. This formula  reduces to that in absence of the strain given in Ref. \cite{xiao07} except for a minus sign because of a sign change in the definition of the Berry curvature. As is known, the orbital magnetic moment for a Bloch wave packet stems from its self rotation  and can be tested in an external magnetic field.

For a zero strain, no charge Hall current is present, as illustrated  in Fig. \ref{CHCVHC}(a). The electron flows from $K$ and $K'$ cones carry  different orbital magnetic moments,  which  gives rise to a finite valley Hall conductivity. The opposite orbital magnetic moment  accumulates on the opposite edges of the Hall bar and may be tested in experiments, e.g. by  giant magneto resistent sensors. For nonzero strain and zero  interaction gap, the net charge Hall current is nonzero since the electrons flow into the same direction with the same orbital magnetic moments in the two cones (see Fig. \ref{CHCVHC}(b)). The interesting point is that the net magnetic moment is non vanishing as a consequence. For the nonzero strain and nonzero  interaction gap, the electron flows may proceed in the same directions; or they may run in opposite directions with different magnitudes of  current density and orbital magnetic moment. Which case is realized in   determined by the relative ratio of the strengthes of the strain and the interaction gap (the latter case is schematically shown in Fig. \ref{CHCVHC}(c)). {The effect could be observed by applying a weak magnetic field in the plane (to minimize the orbital motion and the Zeeman effect could be negligible) to the graphene device with strain or by depositing
 the graphene device on a (magnetic) substrate which breaks the time-reversal symmetry.}

{In summary,  we inspected   the effects of  strain  in graphene within a model that goes beyond the minimal coupling by utilizing
the theory of invariants. We show how the strain tunes  the  topology of the two Dirac cones, the associated  Berry curvatures, the orbital magnetic moments and the conductivities. Unlike the other studies, we also consider the effect of strain under a time-reversal-breaking environment and  point out that a net charge Hall effect could be induced by an off-diagonal strain.
 This is a case which falls in the category of the anomalous Hall effects without spin orbit coupling and could be tested by applying a weak magnetic field in the plane of the graphene device and may be useful in new potential applications in strained-based graphene devices.}


\textbf{Acknowledgements:} We thank R. Winkler and V. K. Dugaev for useful discussions. The work is supported by DFG.

\newpage

\appendix

\section{The Hamiltonian in presence of  a strain  derived from the theory of invariants}
\begin{figure}[tph]
\centering
\includegraphics[width=0.7\textwidth]{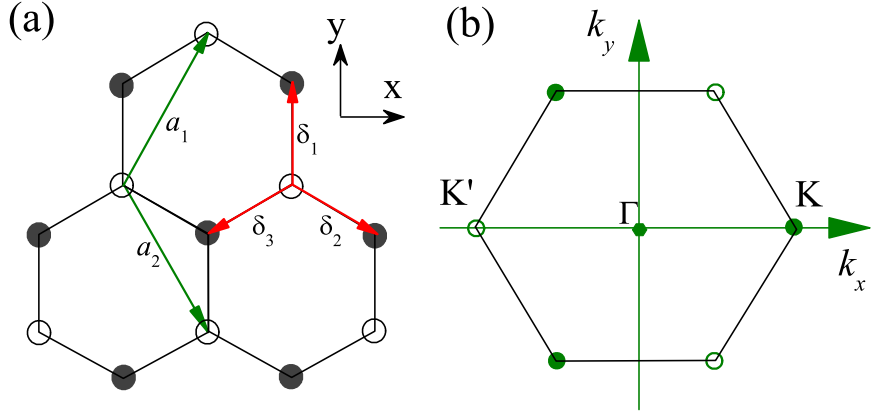}
\caption{(color online) In (a), the crystal structure of graphene in real space is shown with the primitive vectors and the vectors linking the nearest neighbors.  The first Brillouin zone is given in (b).
\label{s1}}
\end{figure}
In Fig. \ref{s1}(a), the graphene lattice structure is shown. $a_{1}$ and $a_{2}$ are the primitive vectors of a unit cell.  Fig. \ref{s1}(b) shows
the corresponding first Brillouin zone. The crystal point group at the $\Gamma$ point is $D_{6h}$ and at the Dirac points, i.e. $K$ and $K'$, is $D_{3h}$. We do not present the detailed group representations here explicitly, for those have been discussed
 in depth in  Refs. \cite{koster,winkler}. The key point is that $D_{3h}$ is a subgroup of $D_{6h}$. Restricting the considerations to
   the low energy regime around the two Dirac points, the constructed Hamiltonian  has to be invariant under the symmetry operations in $D_{3h}$. Therefore, the symmetry operations belonging to $D_{6h}$, but outside $D_{3h}$ will not preserve the invariance of the Hamiltonians at the points $K$ and $K'$. These two Hamiltonians at $K$ and $K'$  are interrelated via the symmetry operations in $D_{6h}$ group which map $K$ onto $K'$ and vice versa. Particularly, we consider one such operation $R_{y}$ which is a reflection plane through $yz$ (and $k_{y}k_{z}$ in the momentum space) perpendicular to the graphene layer. Under this operation, the sublattices are reserved but $K$ is mapped onto $K'$. We then have
\begin{equation}
h_{K'}(\mathbf{k})=\mathcal{D}(R_{y})h_{K}(R^{-1}_{y}\mathbf{k})\mathcal{D}^{-1}(R_{y}),
\label{hkhkp}
\end{equation}
where $\mathcal{D}(R_{y})$ is a representation matrix for the operation $R_{y}$.

Let us define the strain tensor as
\begin{equation}
u_{ij}=\frac{1}{2}\left(\frac{\partial u_{i}}{\partial r_{j}}+\frac{\partial u_{j}}{\partial r_{i}}+\frac{\partial u_{z}}{\partial r_{i}}\frac{\partial u_{z}}{\partial r_{j}}\right),
\end{equation}
where $\mathbf{u}$ is a polar vector indicating the displacements of atoms. In the theory of invariants \cite{toi}, the strain tensor couples with $\mathbf{k}$ to give rise to the irreducible tensors $\mathcal{K}$s under $D_{3h}$ group. We only focus on the $H^{55}$ which is derived by  a multiplication of the irreducible matrices and the irreducible tensors which both belong to the $\Gamma_{5}$ representations. Thus, the Hamiltonian is expressed in terms of $\Gamma_{1}$, $\Gamma_{2}$ and $\Gamma_{6}$, respectively. The corresponding irreducible matrices for these representations are $\mathbf{1}$, $\sigma_{z}$, and $(\sigma_{x},\sigma_{y})$ \cite{winkler}. These four matrices form a complete space in which any $2\times2$ quantity can be expanded in terms of them. The expansion coefficients are just the irreducible tensors $\mathcal{K}$s. In our concrete study about the strain effect, the $\mathbf{d}$ vectors in  Eqs. (1) and (6) in the main text are related to the irreducible tensors, $d_{0}=\mathcal{K}^{1}$, $d_{3}=\mathcal{K}^{2}$, $d_{1}=\mathcal{K}^{6,1}$ and $d_{2}=\mathcal{K}^{6,2}$ where $\mathcal{K}^{\kappa,\lambda}$ indicates irreducible tensors that transform according to the $\kappa$-th irreducible representation of $D_{3h}$. These irreducible tensors except the one for $\Gamma_{1}$ span  the \textit{d} vector space in $K$ cone. Therefore, we can construct a Hamiltonian in the $K$ cone which is invariant upon $D_{3h}$ group
\begin{equation}
h_{K}(\mathbf{k}) =d_{0}(\mathbf{k})+\boldsymbol{\sigma}\cdot\mathbf{d}(\mathbf{k}).
\label{hk}
\end{equation}
Explicitly, the $\mathbf{d}$ vector in the $K$ cone reads
\begin{eqnarray}
d_{1}&=& \tilde{v}_{F}k_{x}+\tilde{\mathbf{v}}\cdot\mathbf{k}-\mathcal{A}_{x}, \notag\\
d_{2} &=& \tilde{v}_{F}k_{y}+(\mathbf{k}\times\tilde{\mathbf{v}})_{z}+\mathcal{A}_{y}, \notag\\
d_{3} &=& \beta(\tilde{v}_{x}k_{y}+\tilde{v}_{y}k_{x}),
\end{eqnarray}
where $\tilde{v}_{F}=v_{F}+b_{62}(u_{xx}+u_{yy})$, $\tilde{v}_{x}=b_{63}(u_{xx}-u_{yy})$, $\tilde{v}_{y}=2b_{63}u_{xy}$, $\mathcal{A}_{x}=b_{61}(u_{xx}-u_{yy})$ and $\mathcal{A}_{y}=b_{61}(2u_{xy})$, and $\beta=b_{21}/b_{63}$ being a dimensionless parameter. The parameters $b$s are material-specific parameters which need to be determined by experiments. The $d_{0}(\mathbf{k})$ will not make a contribution to the Hall effect and Berry's curvature as shown below. However, we write it explicitly here for a reference, i.e. $d_{0}(\mathbf{k})=b_{11}\mathbf{1}+b_{12}(u_{xx}+u_{yy})\mathbf{1}+b_{13}[(u_{yy}-u_{xx})k_{x}+2u_{xy}k_{y}]\mathbf{1}$.
For zero strain, the first term is related to the chemical potential. It will be zero for half-filling.

From Eq. (\ref{hkhkp}) and note $R^{-1}_{y}(\mathcal{A}_{x},\mathcal{A}_{y})=(\mathcal{A}_{x},-\mathcal{A}_{y})$, $R^{-1}_{y}(\tilde{v}_{x},\tilde{v}_{y})=(\tilde{v}_{x},-\tilde{v}_{y})$, we obtain the Hamiltonian in the $K'$ cone
\begin{eqnarray}
h_{K'}(\mathbf{k}) &=& d'_{0}(\mathbf{k})+\boldsymbol{\sigma}\cdot\mathbf{d}'(\mathbf{k}), \notag\\
d'_{1}&=& -\tilde{v}_{F}k_{x}-\tilde{\mathbf{v}}\cdot\mathbf{k}-\mathcal{A}_{x}, \notag\\
d'_{2} &=& \tilde{v}_{F}k_{y}+(\mathbf{k}\times\tilde{\mathbf{v}})_{z}-\mathcal{A}_{y}, \notag\\
d'_{3} &=& d_{3}.
\label{hkp}
\end{eqnarray}
Introducing $\tau_{z}$, the third Pauli matrix, to indicate the valley degree of freedom, its eigenvalue is $\pm1$ for $K(K')$ cone. Therefore, Eqs. (\ref{hk}) and (\ref{hkp}) can be compacted into the Hamiltonian given by Eq. (6) in the main text.

\section{Berry's curvature}
For the Hamiltonian $h_{K}$, we parameterize $\mathbf{d}$ as $\mathbf{d}=d(\cos\varphi\sin\theta,\sin\varphi\sin\theta,\cos\theta)$ (for  $\mathbf{d'}$, we use $d'$, $\varphi'$ and $\theta'$).  The eigen energies are  $E_{\pm}=d_{0}(\mathbf{k})\pm d(\mathbf{k})$ (for $h_{K'}$, $E'_{\pm}=d'_{0}(\mathbf{k})\pm d'(\mathbf{k})$). The eigenvectors for $h_{K}$ are
\begin{eqnarray}
|\Psi_{+K}\rangle=\left(\cos\frac{\theta}{2}, \sin\frac{\theta}{2}e^{i\varphi}\right)^{T}, \notag\\
|\Psi_{-K}\rangle=\left(\sin\frac{\theta}{2}, -\cos\frac{\theta}{2}e^{i\varphi}\right)^{T},
\label{wvk}
\end{eqnarray}
where $\pm$ stand for the bands with the energies $E_{\pm}$. Similar equations for the $K'$ cone can be derived by substituting $\theta',\varphi'$ into the corresponding places in Eq. (\ref{wvk}). The Berry connection (i.e. a $\mathbf{k}$-space vector potential) in the band $\nu$ is given
\begin{equation}
\mathbf{a}_{\nu\tau}(\mathbf{k})=-i\langle\Psi_{\nu\tau}|\nabla_{\mathbf{k}}|\Psi_{\nu\tau}\rangle,
\label{berryconnection}
\end{equation}
where $\tau$ being the valley index. The corresponding Berry curvature (a $\mathbf{k}$-space magnetic field) is
\begin{equation}
\boldsymbol{\Omega}_{\nu\tau}(\mathbf{k})=\nabla_{\mathbf{k}}\times\mathbf{a}_{\nu\tau}=\Omega_{\nu\tau z}(\mathbf{k})\mathbf{e}_{z}.
\label{berrydefine}
\end{equation}
For graphene, there is only the component in $z$ direction. Therefore, we can obtain the Berry curvature by using the eigenvectors  as \cite{nagaosa06}
\begin{equation}
\Omega_{\nu\tau z}(\mathbf{k})=\nu\frac{\sin\theta_{\tau}}{2}\left(\frac{\partial\theta_{\tau}}{\partial k_{x}}\frac{\partial\varphi_{\tau}}{\partial k_{y}}-\frac{\partial\theta_{\tau}}{\partial k_{y}}\frac{\partial\varphi_{\tau}}{\partial k_{x}}\right).
\label{berrycurvaturea}
\end{equation}

\subsection{Zero strain and zero interaction gap}
We have $h_{K(K')}=\pm\sigma_{x}d_{1}+\sigma_{y}d_{2}$, $d_{1}=v_{F}k_{x}=v_{F}k\cos\phi$, $d_{2}=v_{F}k_{y}=v_{F}k\sin\phi$, and $d_{3}=0$. Thus, we infer $\varphi=\phi$ and $\theta=\frac{\pi}{2}$ for $K$ cone. It is straightforward to show that
\begin{equation}
\mathbf{a}_{+K}=\frac{1}{2k}\hat{e}_{\phi}.
\label{apk}
\end{equation}
This is a vector potential of a monopole in the momentum space (note, there is sign difference in Eq. (\ref{apk}) ("-" instead of "+")  to that in  Ref. \cite{fuchs} because of a sign difference in the definition). The Berry curvature is given as $\Omega_{+Kz}=(1/2)\delta^{2}(\mathbf{k})$. The Berry curvature  changes its sign from one band or valley to the other. It is now evident that  there is a singularity at the Dirac points in the momentum space.

\subsection{Zero strain and nonzero interaction gap}
We have $d_{1}=v_{F}k_{x}$, $d_{2}=v_{F}k_{y}$, and $d_{3}=\frac{\Delta}{2}$. Thus, we have $\varphi=\phi$ and $\theta=\tan^{-1}\frac{2v_{F}k}{\Delta}$ for $K$ cone. Straightforward computations deliver for the Berry connection $\mathbf{a}_{+K}=-\frac{1-\cos\theta}{2k}\hat{e}_{\phi}$ and  for the Berry curvature
\begin{equation}
\Omega_{+Kz}=\frac{3\Delta a^{2}t^{2}}{2(\Delta^{2}+3k^{2}a^{2}t^{2})^{3/2}},
\label{omegakz}
\end{equation}
where $v_{F}=\frac{\sqrt{3}}{2}at$ is used. These results match  exactly those  derived in Ref. \cite{xiao07} with a sign change from "+" to "-" because of the difference in the definition (the definition in our paper is chosen so that the results
are consistent (including the signs) with the results of  the Kubo formula).

\subsection{Derivation of  Equation (4)}
For a general 2D Hamiltonian for a Dirac fermion, like Eq. (\ref{hk}), we can calculate the charge Hall conductivity driven by an external electric field. The current density operator for the $K$ cone is $J^{K}_{i}(\mathbf{k})=\frac{\partial h_{K}}{\partial k_{i}}=j_{0i}+j_{\alpha i}\sigma^{\alpha}$, where $\alpha=1,2,3$, $i=x,y$, and $j_{\alpha i}=\frac{\partial d_{\alpha}(\mathbf{k})}{\partial k_{i}}$. From $\tan\theta=\sqrt{d^{2}_{1}+d^{2}_{2}}/d_{3}$, we get $\frac{\partial\theta}{\partial d_{1}}=\cos\varphi\cos\theta/d$, $\frac{\partial\theta}{\partial d_{2}}=\sin\varphi\cos\theta/d$, and $\frac{\partial\theta}{\partial d_{3}}=-\sin\theta/d$. Thus, we infer
\begin{eqnarray}
\frac{\partial\theta}{\partial k_{x}} &=& j_{1x}\frac{\partial\theta}{\partial d_{1}}+j_{2x}\frac{\partial\theta}{\partial d_{2}}+j_{3x}\frac{\partial\theta}{\partial d_{3}} \notag\\
&=& j_{1x}\frac{\cos\varphi\cos\theta}{d}+j_{2x}\frac{\sin\varphi\cos\theta}{d}-j_{3x}\frac{\sin\theta}{d} \notag\\
&=& \frac{1}{d^{2}\sqrt{d^{2}_{1}+d^{2}_{2}}}\left[d_{1}(j_{1x}d_{3}-j_{3x}d_{1})+d_{2}(j_{2x}d_{3}-j_{3x}d_{2})\right] \notag\\
&=& -\frac{1}{d^{2}\sqrt{d^{2}_{1}+d^{2}_{2}}}\left[\mathbf{d}\times(\mathbf{j}_{x}\times\mathbf{d})\right]_{3},
\label{thetakx}
\end{eqnarray}
where $\mathbf{j}_{x(y)}=(j_{1x(y)},j_{2x(y)},j_{3x(y)})$. Similarly, we find
\begin{eqnarray}
\frac{\partial\theta}{\partial k_{y}} &=& -\frac{1}{d^{2}\sqrt{d^{2}_{1}+d^{2}_{2}}}\left[\mathbf{d}\times(\mathbf{j}_{y}\times\mathbf{d})\right]_{3}, \notag\\
\frac{\partial\varphi}{\partial k_{x}} &=& -\frac{1}{d^{2}_{1}+d^{2}_{2}}\left(\mathbf{j}_{x}\times\mathbf{d}\right)_{3}, \notag\\
\frac{\partial\varphi}{\partial k_{y}} &=& -\frac{1}{d^{2}_{1}+d^{2}_{2}}\left(\mathbf{j}_{y}\times\mathbf{d}\right)_{3}.
\label{thetavarphikxky}
\end{eqnarray}
Substituting Eqs. (\ref{thetakx}) and (\ref{thetavarphikxky}) into Eq. (\ref{berrycurvaturea}), we get the Berry curvature
\begin{eqnarray}
\Omega_{\nu Kz} &=& \frac{\nu}{2}\frac{1}{d^{3}(d^{2}_{1}+d^{2}_{2})}\{\left[\mathbf{d}\times(\mathbf{j}_{x}\times\mathbf{d})\right]_{3}\left(\mathbf{j}_{y}\times\mathbf{d}\right)_{3}-
\left[\mathbf{d}\times(\mathbf{j}_{y}\times\mathbf{d})\right]_{3}\left(\mathbf{j}_{x}\times\mathbf{d}\right)_{3} \} \notag\\
&=& \frac{\nu}{2}\frac{1}{d^{3}}[\varepsilon_{\alpha\beta\gamma}j_{\alpha x}j_{\beta y}d_{\gamma}].
\label{berryc1}
\end{eqnarray}
This  is just  Eq. (4) in the main text. A similar equation for the $K'$ cone can be derived by changing the corresponding $\mathbf{d}$ and $\mathbf{j}_{x(y)}$ vectors. This formula for the Berry curvature is  used in the numerical calculations to produce the results
discussed  in the main text.

\section{Anomalous charge Hall conductivity}
\subsection{Anomalous Charge Hall conductivity for one cone}
We first calculate the charge Hall conductivity for the $K$ cone. The Green's function reads
\begin{equation}
G_{K}(\mathbf{k},i\omega_{n})=[i\omega_{n}-h_{K}(\mathbf{k})]^{-1},
\label{greenfunction}
\end{equation}
where $\omega_{n}$ are the  Matsubara frequencies. This Green's function can be further evaluated as
\begin{equation}
G_{K}(\mathbf{k},i\omega_{n})=\sum_{t}\frac{P_{t}}{i\omega_{n}-E^{K}_{t}(\mathbf{k})},
\label{greenfunction1}
\end{equation}
where $t=\pm$ and $P_{\pm}=(1\pm\hat{d}_{\alpha}\sigma^{\alpha})/2$. We use the formula \cite{qi2006}
\begin{eqnarray}
Q^{K}_{ij}(i\nu_{m}) &=& \frac{1}{\beta A_{0}}\sum_{\mathbf{k}n}\text{Tr}\left[J^{K}_{i}G_{K}(\mathbf{k},i(\omega_{n}+\nu_{m}))J^{K}_{j}G_{K}(\mathbf{k},i\omega_{n})\right] \notag\\
&=& \frac{1}{\beta A_{0}}\sum_{\mathbf{k}n}\sum_{s,t=\pm}\frac{\text{Tr}\left[J^{K}_{i}(\mathbf{k})P_{s}(\mathbf{k})J^{K}_{j}(\mathbf{k})P_{t}(\mathbf{k})\right]}
{\left[i(\omega_{n}+\nu_{m})-E^{K}_{s}(\mathbf{k})\right]\left[i\omega_{n}-E^{K}_{t}(\mathbf{k})\right]}.
\label{qij}
\end{eqnarray}
here $\beta=(k_{B}T)^{-1}$ indicates the temperature. In terms of $Q^{K}_{ij}$ we can calculate the charge Hall conductivity
\begin{eqnarray}
\sigma^{K}_{ij} &=& \frac{e^{2}}{\hbar}\lim_{\omega\rightarrow0}\frac{i}{\omega}Q^{K}_{ij}(\omega+i\delta) \notag\\
&=& -\frac{e^{2}}{\hbar}\frac{i}{A_{0}}\sum_{st}\sum_{\mathbf{k}}\frac{\text{Tr}\left[J^{K}_{i}(\mathbf{k})P_{s}(\mathbf{k})J^{K}_{j}(\mathbf{k})P_{t}(\mathbf{k})\right]}
{[E_{t}(\mathbf{k})-E_{s}(\mathbf{k})]^{2}}[n_{t}(\mathbf{k})-n_{s}(\mathbf{k})], \notag\\
&=& -\frac{e^{2}}{\hbar}\frac{i}{A_{0}}\sum_{\mathbf{k}}\frac{[n_{-}(\mathbf{k})-n_{+}(\mathbf{k})]}{4d^{2}(\mathbf{k})}
\left\{\text{Tr}\left(J^{K}_{i}P_{+}J^{K}_{j}P_{-}\right)-\text{Tr}\left(J^{K}_{i}P_{-}J^{K}_{j}P_{+}\right)\right\}, \label{sigmakij}
\end{eqnarray}
where $P^{\dagger}_{\pm}=P_{\pm}$, and $J^{K,\dagger}_{i}=J^{K}_{i}$. Thus
\begin{eqnarray}
\sigma^{K}_{xy} &=& \frac{e^{2}}{\hbar}\frac{i}{4A_{0}}\sum_{\mathbf{k}}\frac{[n_{-}(\mathbf{k})-n_{+}(\mathbf{k})]}{4d^{2}(\mathbf{k})}\times \notag\\
&\text{Tr}&\left[(j_{0x}+j_{\alpha x}\sigma^{\alpha})(1-\hat{d}_{\beta}\sigma^{\beta})(j_{0y}+j_{\mu y}\sigma^{\mu})(1+\hat{d}_{\nu}\sigma^{\nu})  \right.\notag\\
&-& \left.(j_{0x}+j_{\alpha x}\sigma^{\alpha})(1+\hat{d}_{\beta}\sigma^{\beta})(j_{0y}+j_{\mu y}\sigma^{\mu})(1-\hat{d}_{\nu}\sigma^{\nu})\right] \notag\\
&=& \frac{e^{2}}{\hbar}\frac{i}{4A_{0}}\sum_{\mathbf{k}}\frac{[n_{-}(\mathbf{k})-n_{+}(\mathbf{k})]}{2d^{2}(\mathbf{k})}\text{Tr}
\left[(j_{0x}+j_{\alpha x}\sigma^{\alpha})(j_{0y}+j_{\mu y}\sigma^{\mu})\hat{d}_{\nu}\sigma^{\nu}\right. \notag\\
&-& \left.(j_{0x}+j_{\alpha x}\sigma^{\alpha})\hat{d}_{\beta}\sigma^{\beta}(j_{0y}+j_{\mu y}\sigma^{\mu})\right] \notag\\
&=& \frac{e^{2}}{\hbar}\frac{i}{4A_{0}}\sum_{\mathbf{k}}\frac{[n_{-}-n_{+}]}{2d^{2}(\mathbf{k})}\text{Tr}\left[(j_{0x}+j_{\alpha x}\sigma^{\alpha})j_{\mu y}\hat{d}_{\nu}[\sigma^{\mu},\sigma^{\nu}]\right] \notag\\
&=& -\frac{e^{2}}{\hbar}\frac{1}{4A_{0}}\sum_{\mathbf{k}}\frac{[n_{-}-n_{+}]}{d^{2}(\mathbf{k})}\text{Tr}\left[(j_{0x}+j_{\alpha x}\sigma^{\alpha})\varepsilon_{\mu\nu\gamma}j_{\mu y}\hat{d}_{\nu}\sigma^{\gamma}\right] \notag\\
&=& \frac{e^{2}}{\hbar}\frac{1}{A_{0}}\sum_{\mathbf{k}}[n_{+}-n_{-}]\frac{1}{2d^{2}}\varepsilon_{\alpha\beta\gamma}j_{\alpha x}j_{\beta y}\hat{d}_{\gamma}, \notag\\
&=& \frac{e^{2}}{\hbar}\frac{1}{A_{0}}\sum_{\nu,\mathbf{k}}n_{\nu}\Omega_{\nu Kz}(\mathbf{k}),
\label{sigmaxy}
\end{eqnarray}
where $\Omega_{\nu Kz}(\mathbf{k})=\frac{\nu}{2d^{2}}\varepsilon_{\alpha\beta\gamma}j_{\alpha x}j_{\beta y}\hat{d}_{\gamma}$ is the Berry curvature. Eq. (\ref{sigmaxy}) is just the charge Hall conductivity for the $K$ cone, as given by  Eq. (3) in the main text (cf. also Ref. \cite{nagaosa06}).

\subsection{The total charge Hall conductivity and the valley Hall conductivity}
Accounting for  the  valley degrees of freedom, we may utilize  Pauli matrices $\tau$s to characterize  the valley space.
$d_{0}(\mathbf{k})$ and $d'_{0}(\mathbf{k})$ can be put into a matrix in the valley space
\begin{equation}
D_{0}=\left(\begin{array}{cc}
d_{0}(\mathbf{k}) & 0 \\
0 & d'_{0}(\mathbf{k})
\end{array}
\right).
\label{d0}
\end{equation}
The $\mathbf{d}$ vector is now $\mathbf{D}$, and we have
\begin{equation}
H(\mathbf{k})=D_{0}(\mathbf{k})+\boldsymbol{\sigma}\cdot\mathbf{D}(\mathbf{k}),
\label{hvalley}
\end{equation}
where
\begin{eqnarray}
D_{1}&=& (\tilde{v}_{F}k_{x}+\tilde{\mathbf{v}}\cdot\mathbf{k})\tau_{z}-\mathcal{A}_{x}, \notag\\
D_{2} &=& \tilde{v}_{F}k_{y}+(\mathbf{k}\times\tilde{\mathbf{v}})_{z}+\mathcal{A}_{y}\tau_{z}, \notag\\
D_{3} &=& \beta(\tilde{v}_{x}k_{y}+\tilde{v}_{y}k_{x}).
\end{eqnarray}
Therefore, the Hamiltonian in the valley space is now
\begin{equation}
H(\mathbf{k})=\left(\begin{array}{cc}
h_{K}(\mathbf{k}) & 0 \\
0 & h_{K'}(\mathbf{k})
\end{array}
\right).
\label{hmatrix}
\end{equation}
The charge current density operator is
\begin{eqnarray}
J^{C}_{i}(\mathbf{k}) &=& \frac{\partial H(\mathbf{k})}{\partial k_{i}} \notag\\
&=& \left(\begin{array}{cc}
J^{K}_{i} & 0 \\
0 & J^{K'}_{i}
\end{array}
\right).
\label{chargecurrent}
\end{eqnarray}
The valley current density operator is defined as
\begin{eqnarray}
J^{V}_{i}(\mathbf{k}) &=& \frac{1}{2}\left\{\tau_{z},\frac{\partial H(\mathbf{k})}{\partial k_{i}}\right\} \notag\\
&=& \left(\begin{array}{cc}
J^{K}_{i} & 0 \\
0 & -J^{K'}_{i}
\end{array}
\right).
\label{valleycurrent}
\end{eqnarray}
The Green's function is
\begin{equation}
\mathbf{G}(\mathbf{k},i\omega_{n})=\left(\begin{array}{cc}
G_{K}(\mathbf{k},i\omega_{n}) & 0 \\
0 & G_{K'}(\mathbf{k},i\omega_{n})
\end{array}
\right).
\label{gfmatrix}
\end{equation}
From the Kubo formula it follows
\begin{equation}
Q^{C(V)}_{ij}(i\nu_{m})=\frac{1}{\beta A_{0}}\sum_{\mathbf{k},n}\text{Tr}'\left[J^{C(V)}_{i}(\mathbf{k})\mathbf{G}(\mathbf{k},i(\omega_{n}+i\nu_{m}))
J^{C}_{j}(\mathbf{k})\mathbf{G}(\mathbf{k},i\omega_{n})\right],
\label{qcv}
\end{equation}
where the prime in the $\text{Tr}'$ means that we also need to calculate the trace over the valley space in addition to
 the sublattice space. Let us evaluate the trace over the valley space first. This yields
\begin{eqnarray}
Q^{C(V)}_{ij}(i\nu_{m})&=&\frac{1}{\beta A_{0}}\sum_{\mathbf{k},n}\left\{\text{Tr}\left[J^{K}_{i}(\mathbf{k})G_{K}(\mathbf{k},i(\omega_{n}+i\nu_{m}))
J^{K}_{j}(\mathbf{k})G_{K}(\mathbf{k},i\omega_{n})\right] \right.\notag\\
& \pm & \left.\text{Tr}\left[J^{K'}_{i}(\mathbf{k})G_{K'}(\mathbf{k},i(\omega_{n}+i\nu_{m}))
J^{K'}_{j}(\mathbf{k})G_{K'}(\mathbf{k},i\omega_{n})\right]\right\}.
\label{qijpm}
\end{eqnarray}
Therefore, we obtain
\begin{eqnarray}
\sigma^{C(V)}_{xy} &=& \frac{e^{2}}{\hbar}\lim_{\omega\rightarrow 0}\frac{i}{\omega}Q^{C(V)}_{ij}(\omega+i\delta) \notag\\
&=& \sigma^{K}_{xy}\pm\sigma^{K'}_{xy}.
\label{sigmacvxy}
\end{eqnarray}
The "+" indicates the charge Hall conductivity with the contributions from the two cones, and the "-" denotes the valley Hall conductivity.

%



\begin{thebibliography}{}
\bibitem{foot}The sublattice pseudospin, associated with the two distinct sublattices, is less useful because
     it  depends on the carriers' direction of motion.
\end{thebibliography}
\end{document}